\documentclass{elsart1p}

 \usepackage{graphics}
 \usepackage{graphicx}

\usepackage{amstext,amssymb,amsmath,amscd}

 \usepackage{lineno}

\DeclareGraphicsExtensions{.jpg,.mps,.png,.jpg}

\begin{document}

\begin{frontmatter}



\title{SZ and CMB reconstruction using Generalized
Morphological Component Analysis}

\author[1]{Bobin J.}
\author[1]{Moudden Y.}
\author[1,2]{Starck J.-L.}
\author[3]{Fadili J.} 
\author[4]{N.Aghanim}
\address[1]{DAPNIA-SEDI-SAP, Service d'Astrophysique,
CEA/Saclay, 91191 Gif sur Yvette, France.}
\address[2]{Laboratoire APC, 11 place Marcelin Berthelot 75231 Paris Cedex 05, France.}
\address[3]{GREYC CNRS UMR 6072, Image Processing Group, ENSICAEN 14050, Caen Cedex, France.}
\address[4]{IAS, CNRS \& Univ. Paris Sud, 
B\^at. 121, 91405 ORSAY CEDEX}

\begin{abstract}
In the last decade, the study of cosmic microwave background
(CMB) data has become one of the most powerful tools to study and
  understand the Universe. More precisely,
measuring the CMB power spectrum leads to the estimation of most
cosmological parameters. Nevertheless, accessing such precious
physical information requires extracting several different
astrophysical components from the data. Recovering those astrophysical
sources (CMB, Sunyaev-Zel'dovich clusters, galactic dust) thus amounts
to a component separation problem which has already led to an intense
activity in the field of CMB studies. In this paper, we introduce a
new sparsity-based component separation method coined Generalized
Morphological Component Analysis (GMCA). The GMCA approach is
formulated in a Bayesian {\em maximum a posteriori} (MAP)
framework. Numerical results show that this new source recovery
technique performs well compared to state-of-the-art component
separation methods already applied to CMB data.
\end{abstract}

\end{frontmatter}

\section*{Introduction}
\label{sec:intro}
Investigating Cosmic Microwave Background (CMB) data is of huge
scientific importance as it improves our knowledge of the Universe
\cite{Jungman}. Indeed, most cosmological parameters can be
derived from the study of CMB data. In the last decade several
experiments (Archeops, Boomerang, Maxima, WMAP - \cite{WMAP}) have
already provided large amounts of data and astrophysical information. The
forthcoming Planck ESA mission will provide new accurate data
requiring effective data analysis tools. More precisely, recovering
useful scientific information requires disentangling in the CMB data
the contribution of several astrophysical components namely CMB itself, Galactic emissions from dust and synchrotron,
Sunyaev-Zel'dovich (SZ) clusters \cite{sz} to name a few.  In the
frequency range used for CMB observations \cite{Bouchet}, the observed
data combines contributions from distinct astrophysical components the
recovery of which falls in the frame of component separation.\\
Following a standard practice in the field of component or
source separation, which has physical grounds here, the observed sky
is modeled as a linear mixture of statistically independent
components.  The observation with detector $i$ is then a noisy linear
mixture of $n$ independent sources $\{s_j\}_{j=1,\cdots,n}$ : $x_i =
\sum_{j=1}^n a_{ij} s_j + n_i$. The coefficient $a_{ij}$ reflects the
emission law of source $s_j$ in the frequency band of the $i$-th
sensor; $n_i$ models instrumental noise. When $m$ sensors provide
observations at different frequencies, this linear mixture model can
be rewritten in a more convenient matrix formulation :
\begin{equation}
\label{eq:lm_model}
{\bf X} = {\bf AS} + {\bf N}
\end{equation}
where ${\bf X}$ is the $m \times t$ data matrix the rows of which are
the observed data maps in each channel, ${\bf A}$ is the $m \times n$
mixing matrix, ${\bf S}$ is the $n \times t$ source matrix the rows of
which are the sources $s_j$, and ${\bf N}$ is the $m \times t$ noise
matrix. In practice, both the sources ${\bf S}$ and their emission
laws
${\bf A}$ may be unknown or only partly known. A
component separation technique then aims at estimating both ${\bf S}$
and ${\bf A}$ from the data ${\bf X}$. This problem refers to Blind
Source Separation (BSS).\\ 
Amongst all the physical components mixed in the observed data, each
one raises scientific interest. Thus it would be worthwhile to devise
a separation technique able to differentiate effectively between most
physical components. Up to now, several source separation techniques
have already been used in the field of CMB data studies. In this
paper, we concentrate on two particular components: the CMB and the SZ
components. For such processes, state-of-the-art blind separation methods
used on CMB data are:
\begin{itemize}
\item JADE which is a classical Independent Component Analysis
  technique based on fourth order statistics. Its effectiveness at
  extracting non-Gaussian components such as the SZ map was shown in 
  \cite{Pires}.
\item Spectral Matching ICA (SMICA) (see \cite{Delab} and
\cite{YM}) has been devised to accurately separate the CMB
component. SMICA assumes the case of mixed stationary Gaussian
components in a noisy environment. It is based on second order
statistics. In the Fourier representation, colored stationary Gaussian
components are discernible based on the diversity of their power
spectra. SMICA is then well adapted to Gaussian components such as
CMB.
\end{itemize}

\noindent
Neither of the aforementioned techniques 
is able to effectively
extract both the SZ and CMB maps. In this paper we propose a novel
sparsity-based component separation technique coined Generalized
Morphological Component Analysis (GMCA) which turns out to be well
suited 
for the recovery of CMB and SZ
components. Section~\ref{sec:gmca} describes the GMCA model and the
algorithm proposed to solve the corresponding optimization
problem. Numerical experiments are given which illustrate the
astounding performances of GMCA for CMB and SZ extraction in
Section~\ref{sec:results}. Finally, we show in Section~\ref{sec:pbss}
that GMCA is versatile enough to account for physical priors.

\section{Generalized Morphological Component Analysis}
\label{sec:gmca}
\subsection{The GMCA model}
\label{sec:model}
In the previous section, we introduced the linear mixture model in
Equation~\ref{eq:lm_model}. We further assume that all the
protagonists of the
model in Equation~\ref{eq:lm_model} are random components (variables
or vectors). More particularly, the entries of the noise matrix ${\bf
N}$ are assumed to be \textit{independently} distributed according to
a zero mean Gaussian distribution with variance $\sigma_i^2$ depending
on the detector. From physical considerations, ${\bf N}$ models
instrumental noise the level of which varies independently from one
detector to another. ${\bf N}$ is thus a random Gaussian variable with
zero mean and covariance matrix ${\bf \Gamma_N} =
\mbox{diag}(\sigma_1^2,\cdots,\sigma_m^2)$. In practice, as the
detectors are assumed to be accurately calibrated, ${\bf \Gamma_N}$ is
known with high precision. The log-likelihood function is then the
following one :
\begin{equation}
\label{eq:ll}
\log P({\bf X} \big| {\bf A},{\bf S},{\bf \Gamma_N}) = -\frac{1}{2} \|{\bf X} - {\bf AS}\|_{2,{\bf \Gamma_N}}^2 + C
\end{equation}
where $C$ is a constant. The notation $\| . \|_{2,{\bf \Gamma_N}}^2$
stands for the Frobenius norm of ${\bf Y}$ in the noise covariance
metric : $\| Y \|_{2,{\bf \Gamma_N}}^2 = \mbox{ Trace}\left( {\bf Y}^T
{\bf \Gamma_N}^{-1} {\bf Y}\right)$. From a Bayesian point of view,
adding physical priors should help the separation task. We first
assume no particular knowledge about the emission laws of the
components modeled by ${\bf A}$. For simplicity, we consider that each
entry of the mixing matrix ${\bf A}$ is
\textit{i.i.d.}\footnote{Independently and identically distributed.}
from a uniform zero mean distribution. Note that it would be possible
to add some physical constraint on the emission laws reflected in
${\bf A}$.\\ In the general case, source separation is merely a
question of diversity and contrast between the sources (see
\cite{Cardo1}). For instance, on the one hand JADE relies on
non-Gaussianity to distinguish between the sources. On the other,
SMICA takes advantage of the diversity of the mixed components' power
spectra to achieve the separation task. ``Non-Gaussianity" and ``power
spectra diversity" are contrasts between the sources. A
combination of both characteristics, ``Non-Gaussianity" and ``power
spectra diversity", was also proposed to separate CMB from kinetic SZ
signal which are otherwise undistinguishable \cite{forni}.  Recent
work has already emphasized on sparsity as a source of diversity to
improve component separation (see \cite{Zibu} and
\cite{MMCA}). In that setting, each source $\{s_j\}_{j=1,\cdots,n}$ is
assumed to be sparse in a representation (potentially overcomplete)
$\mathcal{D}$. Formally, $\mathcal{D}$ is a fixed dictionary of signal
waveforms written as a $T \times t$ matrix. We define the set of
projection coefficients $\alpha_j$ such that : $\forall j \in
\{1,\cdots,n\}, \quad s_j = \alpha_j \mathcal{D}$. Any source $s_j$ is
said to be sparse in $\mathcal{D}$ if most of the entries of
$\alpha_j$ are nearly zero and only a few have ``significant"
amplitudes. When $\mathcal{D}$ is overcomplete ($T > t$),
$\mathcal{D}$ is called a dictionary. 
The attractiveness of overcomplete representations in image processing theory resides in the potentially very sparse representations that they make possible (see e.g. \cite{DH} and references therein). 
In the field of basic source separation we showed in \cite{MMCA} that
morphological diversity and sparsity are key properties leading to
better separation. We noticed that the gist of sparsity-based source
separation methods leans on the rationale : ``\textit{independent
sources are distinctly sparse in a dictionary $\mathcal{D}$}". In that
study, we considered the simple case of morphologically different
sources : components were assumed to be sparsely represented in
different sub-dictionaries. We illustrated that such a  
sparsity based prior provides a very effective way to distinguish between sources. In the
present paper, we focus on a more general setting : the sources can
have similar morphologies (\textit{i.e.} all the sources are sparsely
represented over the whole $\mathcal{D}$). When the overcomplete
dictionary $\mathcal{D}$ is made of the union of $D$ orthonormal bases
(\textit{i.e.} $\mathcal{D} = \left[\Phi_1,\cdots,\Phi_D\right]$) then
each source is modeled as the linear combination of $D$ so-called
morphological components (see \cite{SED} for details on Morphological
Component Analysis) - each morphological component being sparse in a
different orthonormal basis $\{\Phi_1,\cdots,\Phi_D\}$:
\begin{equation}
\forall  j\in \{1,\cdots,n\}, \quad s_j   =   \sum_{k=1}^D \varphi_{jk} = \sum_{k=1}^D \alpha_{jk} \Phi_k
\end{equation}
From a statistical viewpoint, we assume that the entries of
$\alpha_{jk} = \varphi_{jk}\Phi_k^T$ are \textit{i.i.d.} from a
Laplacian probability distribution with scale parameter $1/\mu$:
\begin{equation}
\label{eq:source_prior}
P(\varphi_{jk}) \propto \exp\left(- \mu \|\varphi_{jk}\Phi_k^T\|_1\right)
\end{equation}
where the $\ell_1$-norm $\|.\|_1$ stands for $\|x\|_1 = \sum_{p=1}^t
|x[p]|$ in which $x[p]$ is the $p$-th entry of $x$. In practice, the
Laplacian prior is well adapted to model leptokurtic sparse
signals. We classically assume that the morphological components are
statistically mutually independent : $P({\bf S}) = \prod_{j,k}
P(\varphi_{jk})$. Estimating the sources ${\bf S}$ is then equivalent to
estimating the set of morphological components
$\{\varphi_{jk}\}_{j=1,\cdots,n;k=1,\cdots,D}$. In this Bayesian
context, we propose to estimate those morphological components
$\{\varphi_{jk}\}$ and the mixing matrix ${\bf A}$ from a
\textit{maximum a posteriori} (MAP) leading to the following
optimization problem:
\begin{equation}
\left\{\{\hat{\varphi}_{jk}\},{\bf \hat{A}}\right\} = \arg\max_{\{\varphi_{jk}\},{\bf A}} P({\bf X}|{\bf A},\{\varphi_{jk}\},{\bf \Gamma_N}) \prod_{j,k}P(\varphi_{jk}) P({\bf A})
\end{equation}
where we further assumed that the morphological components
$\{\varphi_{jk}\}$ are independent of ${\bf A}$. Owing to
Equations~\ref{eq:ll} and \ref{eq:source_prior}, the mixing matrix
${\bf A}$ and the morphological components $\{\varphi_{jk}\}$ are
obtained by minimizing the following negative log \textit{a
posteriori}:
\begin{equation}
\label{eq:optim}
\left\{\{\hat{\varphi}_{jk}\},{\bf \hat{A}}\right\} = \arg\min_{\{\varphi_{jk}\},{\bf A}}\|{\bf X} - {\bf AS}\|_{2,{\bf \Gamma_N}}^2 + 2 \mu \sum_{j=1}^n \sum_{k=1}^D \|\varphi_{jk}\Phi_k^T\|_1
\end{equation}
where $\forall j \in \{1,\cdots,n\},\quad s_j = \sum_{k=1}^D
\varphi_{jk}$. Equation~\ref{eq:optim} leads to the GMCA estimates of
the sources and the mixing matrix in a general sparse component
separation context. Interestingly, in the case of CMB data, the
sources we look for (CMB, galactic dust and SZ) are quite sparse in
the same unique orthonormal wavelet basis. The dictionary
$\mathcal{D}$ then reduces to a single orthonormal basis $\Phi$. In
that case, since $\Phi$ is unitary, Equation~\ref{eq:optim} can be
rewritten as follows :
\begin{eqnarray}
\label{eq:foptim}
\left\{{\bf \hat{\alpha}},{\bf \hat{A}}\right\} &=& \arg\min_{{\boldsymbol \alpha},{\bf A}} \|{\bf X}\Phi^T - {\bf A\alpha}\|_{2,{\bf \Gamma_N}}^2 + 2 \mu \|{\boldsymbol \alpha}\|_1 \nonumber \\
						&=& \arg\min_{{\boldsymbol \alpha},{\bf A}} f_\mu({\boldsymbol \alpha},{\bf A}) = {\arg\min}_{{\boldsymbol \alpha},{\bf A}} f_0({\bf A}, {\boldsymbol \alpha}) + 2\mu f_1({\boldsymbol \alpha})
\end{eqnarray}
where ${\boldsymbol \alpha} = {\bf S}\Phi^T$. Note that the estimation
is done in the sparse representation $\Phi$ requiring a single
transform of the data ${\bf X}\Phi^T$. To remain computationally
efficient, GMCA relies on practical transforms which generally involve
fast implicit operators (typical complexity of
$\mathcal{O}\left(t\right)$ or $\mathcal{O}\left(t \log t
\right)$). In \cite{Zibu}, the authors also used a unique orthonormal
wavelet basis. While a gradient descent is used in \cite{Zibu}, we use
a fast and efficient iterative thresholding optimization scheme which
we describe in the next section.

\subsection{Solving the optimization problem}
\label{sec:algo}
The \textit{maximum a posteriori} estimates of the coefficients
${\boldsymbol \alpha}$ and the mixing matrix in
Equation~\ref{eq:foptim} lead to a non-convex minimization
problem. Note that in Equation~\ref{eq:foptim} the functional to be
minimized suffers from several invariances : any permutation or
rescaling of the sources and the mixing matrix leaves the product $\bf
A{\boldsymbol \alpha}$ unaltered. The scale invariance is
computationally alleviated by forcing the columns of ${\bf A}$ to have
unit $\ell_2$ norm~: $\forall i\in{\{1,\cdots,n\} },\quad a^{i^T}a^i = 1$
where $a^i$ is the $i$-th column of ${\bf A}$.\\ 
As solutions of problem~(\ref{eq:foptim}) have no explicit formulation, we propose
solving it by means of a block-coordinate relaxation iterative
algorithm such that each iteration $(h)$ is decomposed into two steps
: (i) estimation of the sources ${\bf S}$ assuming the mixing matrix
is fixed to its current estimate ${\bf \hat{A}}^{(h-1)}$ and (ii)
estimation of the mixing matrix assuming the sources are fixed to
their current estimates ${\bf \hat{S}}^{(h)}$. It is not difficult to
see that the objective MAP functional in (\ref{eq:foptim}) is
continuous on its effective domain and has compact level
sets. Moreover, this objective function is convex in the source
coefficient vectors $(\alpha_1,\ldots,\alpha_n)$, and $f_0$ has an
open domain, is continuous and G\^ateaux differentiable. Thus by
\cite[Theorem 4.1]{Tseng2001}, the iterates generated
by our alternating algorithm are defined and bounded, and each
accumulation point is a stationary point of the MAP functional. In
other words, our iterative algorithm will converge. Hence, at
iteration $(h)$, the sources are estimated from a \textit{maximum a
posteriori} assuming ${\bf A} = {\bf \hat{A}}^{(h-1)}$. By classical
ideas in convex analysis, a necessary condition for ${\boldsymbol
\alpha}$ to be a minimizer is that the zero is an element of the
subdifferential of the objective at ${\boldsymbol \alpha}$. We
calculate\footnote{For clarity, we drop the superscript $(h-1)$ and
write $\hat{\bf A} = \hat{\bf A}^{(h-1)}$.}:
\begin{equation}
\label{eq:subdiff}
\partial_{\boldsymbol \alpha} f_\mu({\boldsymbol \alpha},{\bf A})= -2{\bf {\bf A}}^T{\bf \Gamma_N}^{-1}({\bf X}\Phi^T - {\bf A}{\boldsymbol \alpha}) + 2\mu \partial_{\boldsymbol \alpha} \|{\boldsymbol \alpha}\|_1
\end{equation}
where $\partial_{\boldsymbol \alpha} \|{\boldsymbol \alpha}\|_1$ is
defined as (owing to the separability of the prior):
\[
\partial_{\boldsymbol \alpha} \|{\boldsymbol \alpha}\|_1 = \left\{U \in \mathbb{R}^{n \times t} \Bigg| 
\begin{array}{ccc}
U{j,k} & = \mbox{ sign}(\alpha_{j,k}), & ~ \alpha_{j,k} \neq 0 \\
U{j,k} & \in [-1,1], & ~ \alpha_{j,k} = 0
\end{array} \right\}.
\]
Hence, Equation \ref{eq:subdiff} can be rewritten equivalently as two
conditions leading to the following (proximal) fixed point equation:
\begin{equation}
\label{eq:it_estim1}
\begin{array}{cc}
\hat{\alpha}_{j,k} = 0, & \text{if} ~ \left|{\left({\bf A}^T{\bf \Gamma_N}^{-1}{\bf X}\Phi^T\right)}_{j,k} \right| \leq \mu \\
{\bf {\bf A}}^T{\bf \Gamma_N}^{-1}({\bf X}\Phi^T - {\bf A}\hat{\boldsymbol \alpha}) = \mu \mbox{ sign}\left(\hat{\boldsymbol \alpha}\right), & \text{otherwise}. 
\end{array}
\end{equation}
Unfortunately, Equation~\ref{eq:it_estim1} has no closed-form solution
in general. It must be iterated and is thus computationally
demanding. Fortunately, it can be simplified when ${\bf A}$ has nearly
orthogonal columns in the noise covariance matrix (\textit{i.e.} ${\bf
\hat{A}}^T{\bf \Gamma_N}^{-1}{\bf \hat{A}} \simeq
\mbox{diag}\left({\bf \hat{A}}^T{\bf \Gamma_N}^{-1}{\bf
\hat{A}}\right)$). Let ${\bf C} = {\left({\bf \hat{A}}^T{\bf
\Gamma_N}^{-1}{\bf \hat{A}}\right)}^{-1}{\bf A}^T{\bf
\Gamma_N}^{-1}{\bf X}\Phi^T$, Equation~\ref{eq:it_estim1} boils down
to the following set of equations $\forall j\in\{1,\cdots,n\}$:
\begin{equation}
\label{eq:it_estim}
\begin{array}{ccc}
\hat{\alpha}_{j,k} & = 0, \quad \text{if} ~ \left|{{\bf C}}_{j,k} \right| \leq \mu^{(h)} \sigma_j^2 \\
\hat{\alpha}_j & = {\left[{\bf C}\right]}_j - \mu \sigma_j^2  \mbox{ sign}\left(\hat{\alpha}_j\right), \quad \text{otherwise}.
\end{array}
\end{equation}
where $[{\bf Y}]_j$ is the $j$-th row of ${\bf Y}$. In practice, even if
the approximation we make is not strictly valid, such a simplification
leads to good computational results. These equations are known as
soft-thresholding with threshold $\mu^{(h)} \sigma_j^2$. We define
$\mathrm{ST}_{\delta}(.)$, the soft-thresholding operator with
threshold $\delta$. At iteration $(h)$, the sources are thus estimated
such that:
\begin{equation}
\hat{\alpha}_j^{(h)} = \mathrm{ST}_{\mu^{(h)} \sigma_j^2}\left(\left[{\bf C}\right]_j\right)
\end{equation}
The $j$th source is reconstructed as $\hat{s}_j^{(h)} =
\hat{\alpha}_j^{(h)}\Phi$. The mixing matrix ${\bf A}$ is then estimated
by a maximum likelihood estimate amounting to a simple least-squares
update assuming ${\bf S}$ is fixed. The GMCA algorithm is then described
as follows :
\begin{flushleft}
\vspace{0.15in}
\centering
\begin{tabular}{|c|} \hline
\begin{minipage}[h]{0.95\linewidth}
\vspace{0.025in} \footnotesize{\textsf{1. Set the number of iterations
$I_{\max}$ and thresholds $\delta_j^{(0)} = \mu^{(0)}\sigma_j^2$\\}
\textsf{2. While each $\mu^{(h)}$ is higher than a given lower bound
$\mu_{min}$ (e.g. can depend on the noise variance), \\}
\hspace{0.1in} \textsf{-- Proceed with the following iteration to
estimate source coefficients ${\boldsymbol \alpha}$ at iteration $h$
assuming ${\bf A}$ is fixed:}
\hspace{0.2in} \textsf{$\hat{\alpha}_j^{(h)} = \mathrm{ST}_{\mu^{(h)} \sigma_j^2}\left(\left[{\left({\bf \hat{A}}^T{\bf \Gamma_N}^{-1}{\bf \hat{A}}\right)}^{-1}{\bf \hat{A}}^T{\bf \Gamma_N}^{-1}{\bf X}\Phi^T\right]_j\right)$:\\}
\hspace{0.1in} \textsf{-- Update $\bf A$ assuming ${\boldsymbol \alpha}$ is fixed :}
\hspace{0.2in} \textsf{${\bf \hat{A}}^{(h)} = {\bf X}\Phi^T{\bf \hat{\boldsymbol \alpha}}^T\left({\bf \hat{\boldsymbol \alpha}}{\bf \hat{\boldsymbol \alpha}}^T\right)^{-1}$\\}
\textsf{-- Decrease the threshold $\mu^{(h)}$ following a given strategy}}
\vspace{0.05in}
\end{minipage}
\\\hline
\end{tabular}
\vspace{0.15in}
\end{flushleft}
Note that the overall optimization scheme is based on an iterative and
alternate thresholding algorithm involving a \textit{coarse to fine}
estimation process. Indeed, \textit{coarse} versions of the sources
(\textit{i.e.} containing the most ``significant" features of the
sources) are first computed with high values of $\mu^{(h)}$.
In the early stages of the algorithm, the mixing matrix is then
estimated from the most ``significant" features of the sources which
are less perturbed by noise. The estimation of ${\bf A}$ and ${\bf S}$ is
then refined at each iteration as $\mu^{(h)}$ (and thus the thresholds
$\{\mu^{(h)}\sigma_j^2\}_{j=1,\cdots,n}$) decreases towards a final
value $\mu_{min}$. We already used this minimization scheme in
\cite{MMCA} where this optimization process provided robustness and
helped convergence even in a noisy context. Experiments in
Section~\ref{sec:results} illustrate that it achieves good results
with GMCA as well.

\section{Application to CMB and SZ reconstruction}
\label{sec:results}
\subsection{Blind component separation}
\label{sec:bss}
\begin{figure}[h]
\begin{minipage}[b]{0.3\linewidth}
    \centerline{\includegraphics[width=3cm]{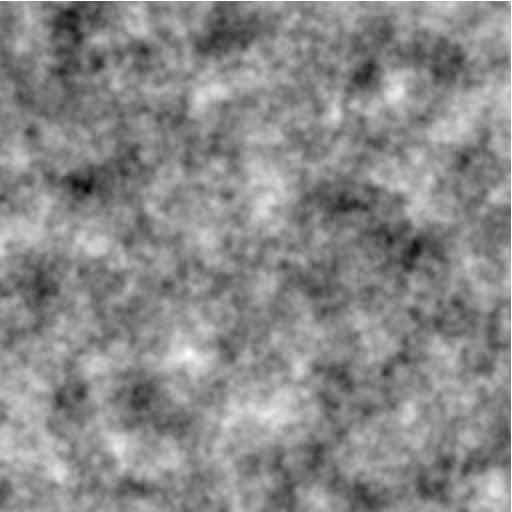}}
\end{minipage}
\hfill
\begin{minipage}[b]{0.3\linewidth}
    \centerline{\includegraphics[width=3cm]{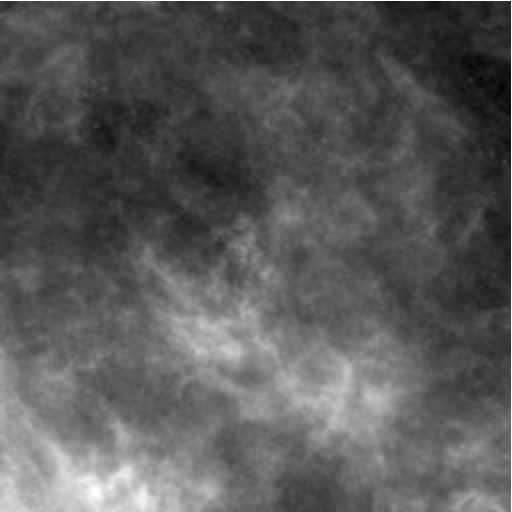}}
\end{minipage}
\hfill
\begin{minipage}[b]{0.3\linewidth}
    \centerline{\includegraphics[width=3cm]{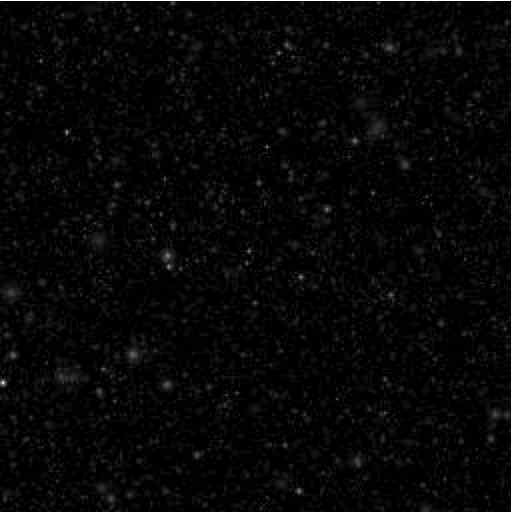}}
\end{minipage}
\vspace{-0.1in} \caption{\textbf{The simulated sources - Left: }
CMB. \textbf{Middle: } galactic dust emission. \textbf{Right: } SZ map.}
\label{fig:sources}
\end{figure}

\begin{figure}[htb]
\begin{minipage}[b]{0.3\linewidth}
    \centerline{\includegraphics[width=3cm]{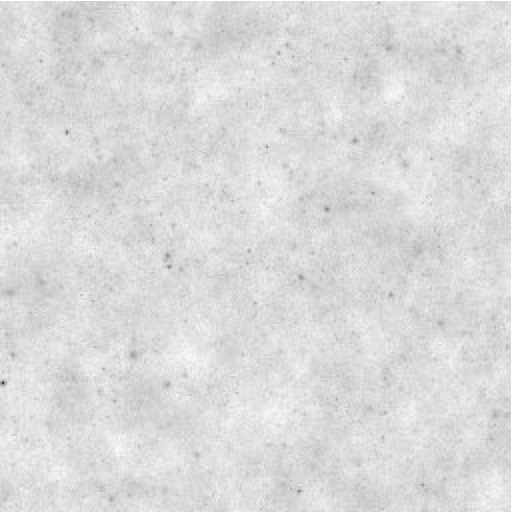}}
\end{minipage}
\hfill
\begin{minipage}[b]{0.3\linewidth}
    \centerline{\includegraphics[width=3cm]{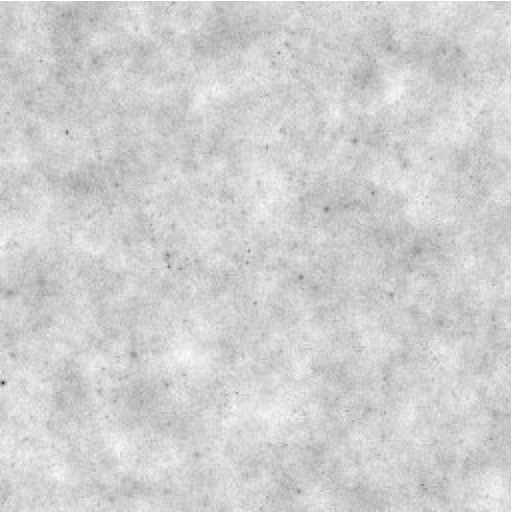}}
\end{minipage}
\hfill
\begin{minipage}[b]{0.3\linewidth}
    \centerline{\includegraphics[width=3cm]{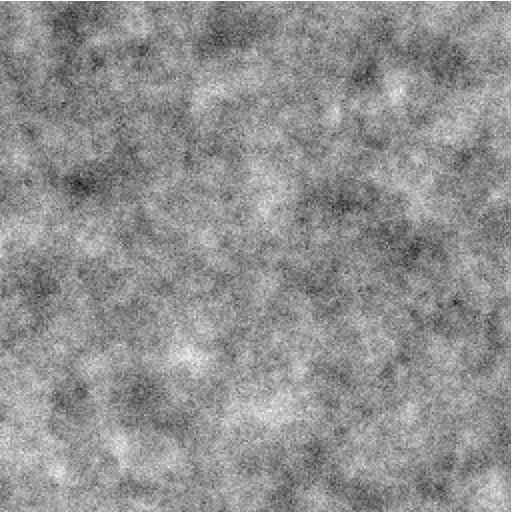}}
\end{minipage}
\vspace{0.1in}
\vfill
\begin{minipage}[b]{0.3\linewidth}
    \centerline{\includegraphics[width=3cm]{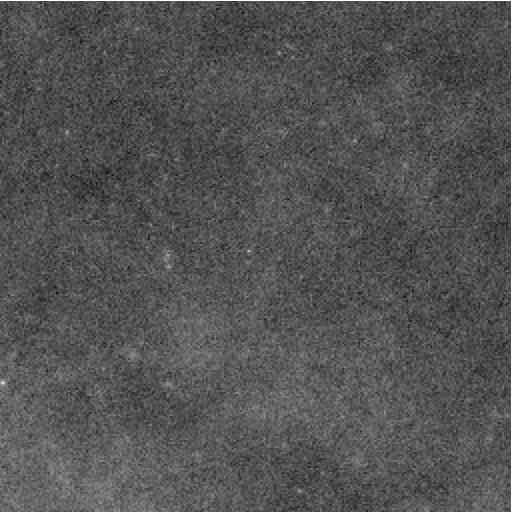}}
\end{minipage}
\hfill
\begin{minipage}[b]{0.3\linewidth}
    \centerline{\includegraphics[width=3cm]{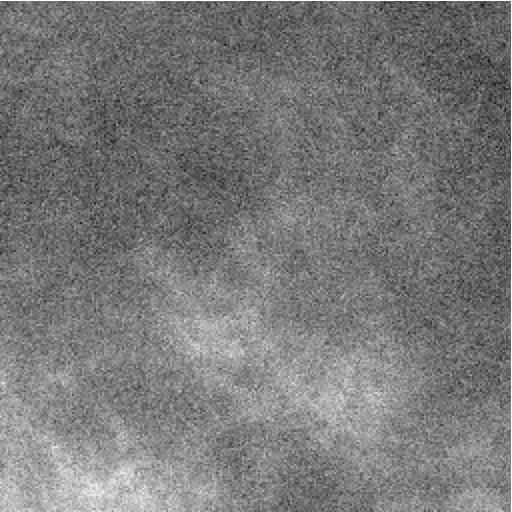}}
\end{minipage}
\hfill
\begin{minipage}[b]{0.3\linewidth}
    \centerline{\includegraphics[width=3cm]{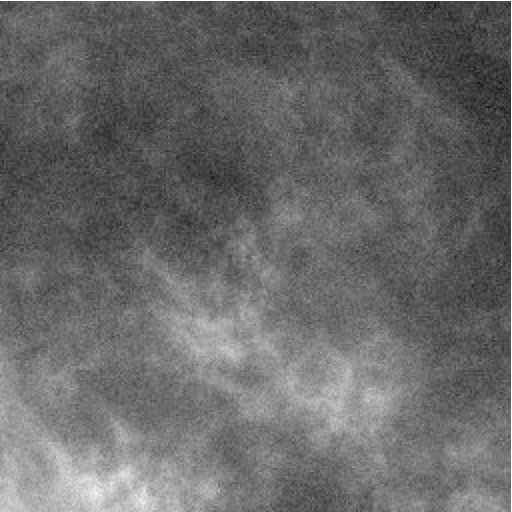}}
\end{minipage}
\vspace{-0.1in} \caption{\textbf{The observed CMB data - global SNR =
2.7dB } } \label{fig:mixtures}
\end{figure}

\begin{figure}[htb]
\vfill
\hspace{0.1in}
\begin{minipage}[b]{0.5\linewidth}
    \centerline{\includegraphics[width=7cm]{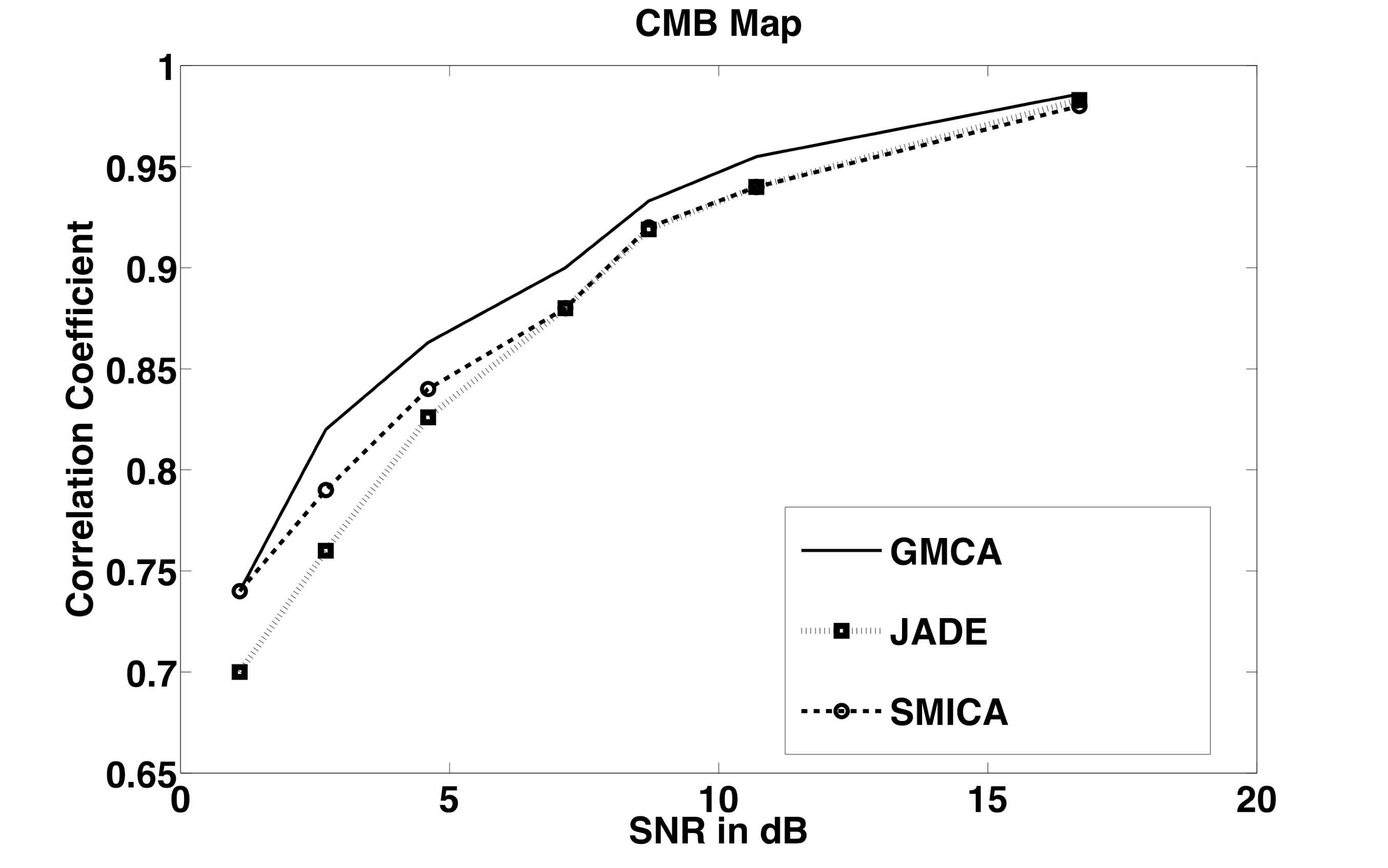}}
\end{minipage}
\hfill
\begin{minipage}[b]{0.5\linewidth}
    \centerline{\includegraphics[width=7cm]{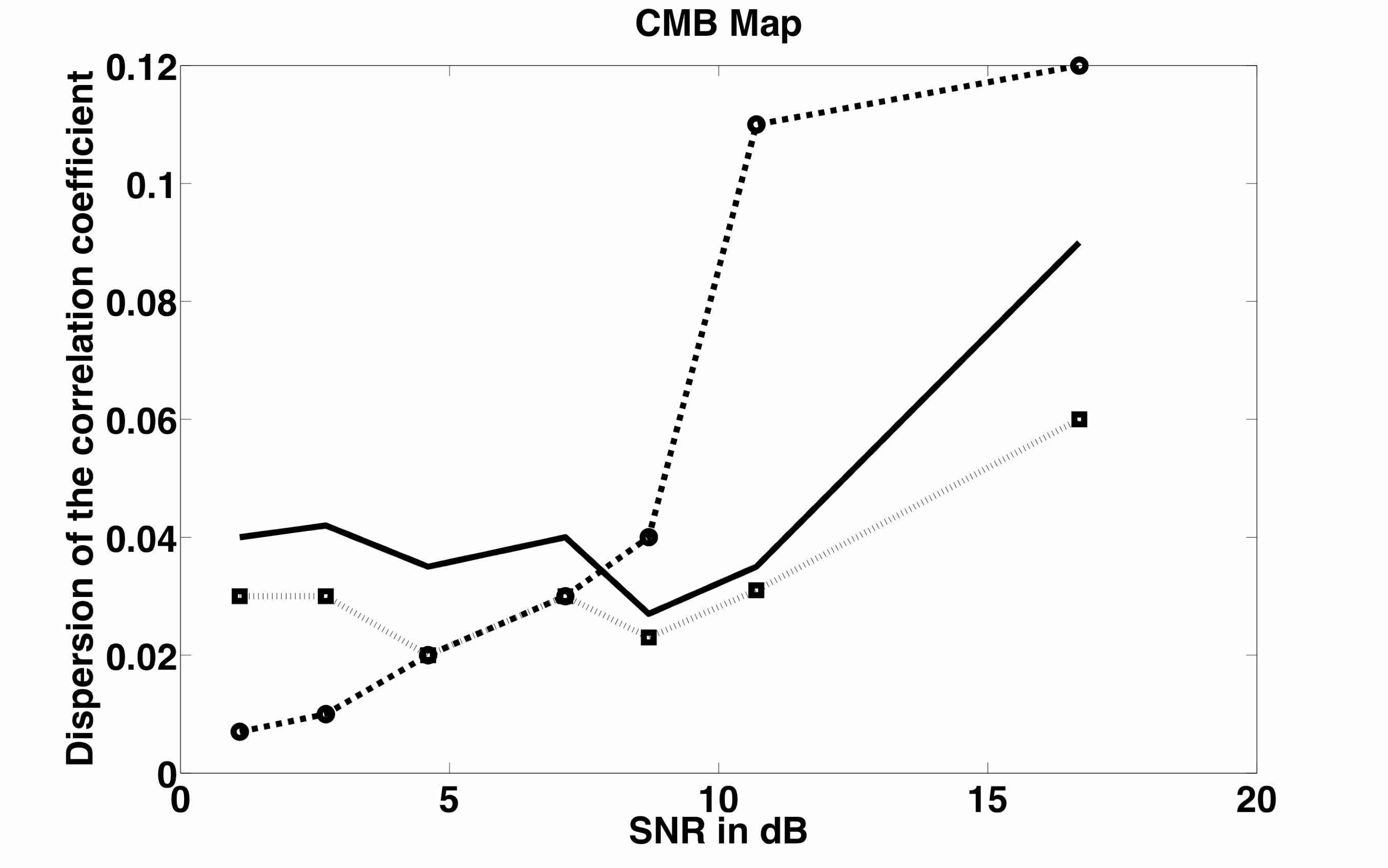}}
\end{minipage}
\vfill
\hspace{0.1in}
\begin{minipage}[b]{0,5\linewidth}
    \centerline{\includegraphics[width=7cm]{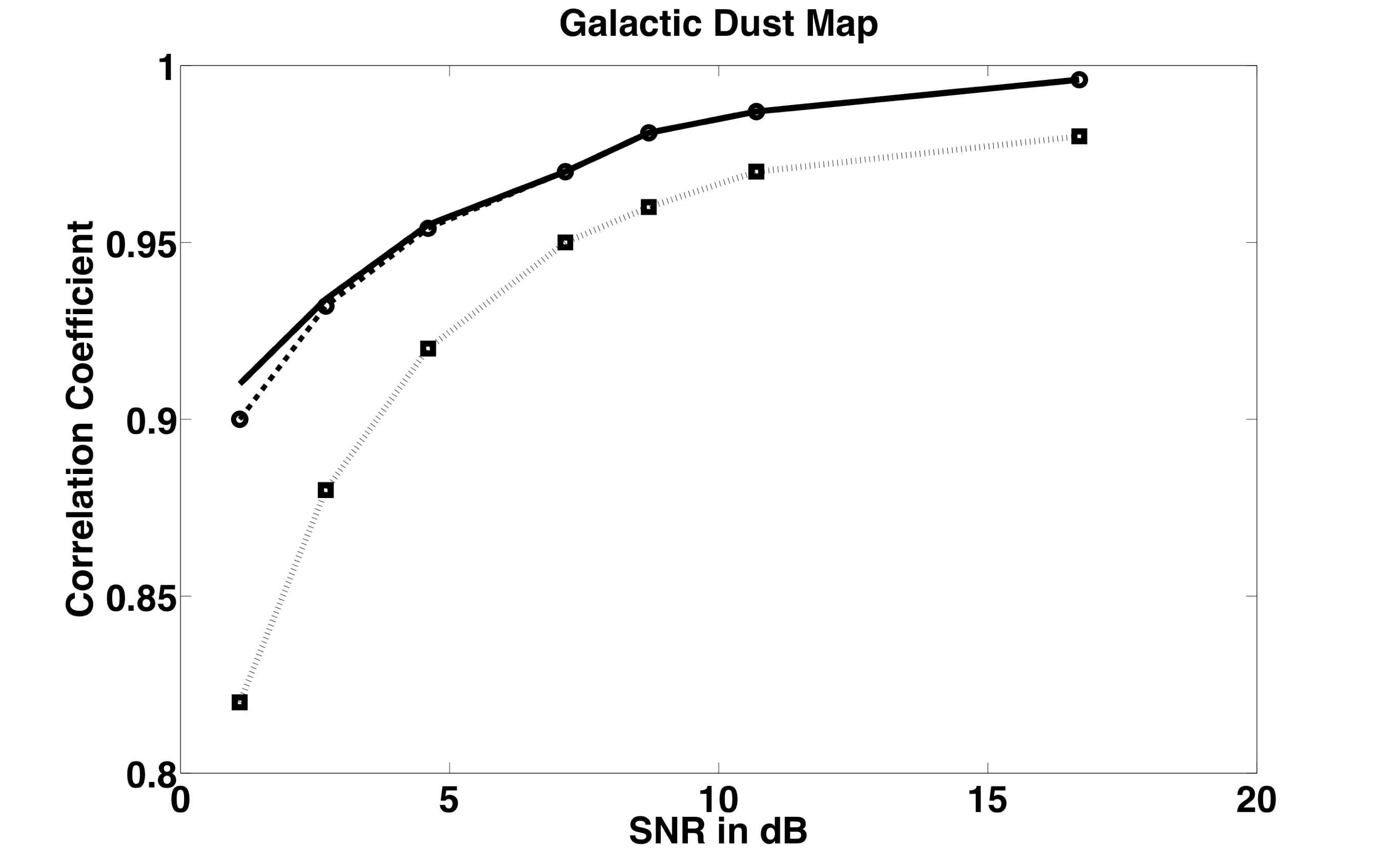}}
\end{minipage}
\hfill
\begin{minipage}[b]{0,5\linewidth}
    \centerline{\includegraphics[width=7cm]{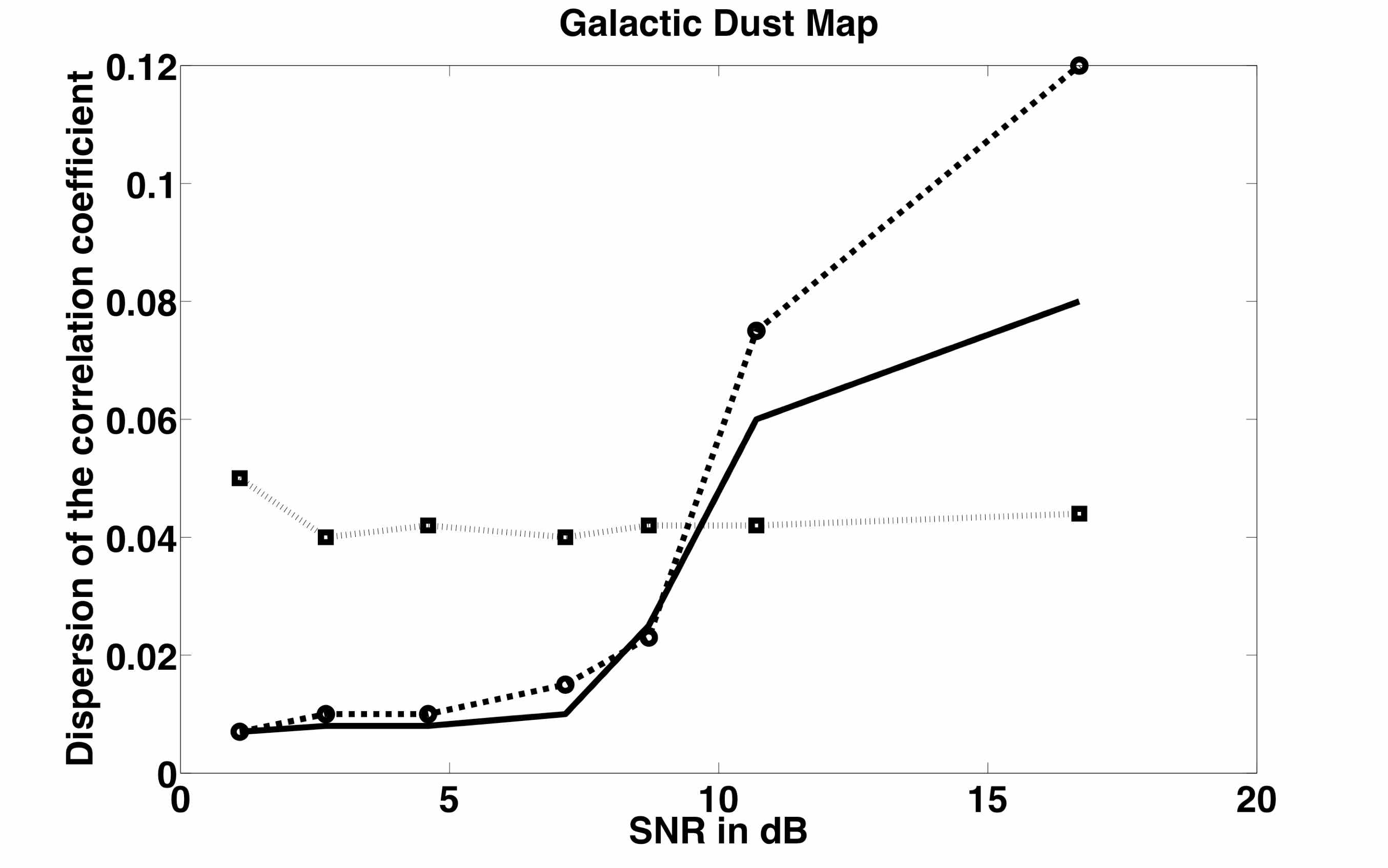}}
\end{minipage}
\vfill
\hspace{0.1in}
\begin{minipage}[b]{0,5\linewidth}
    \centerline{\includegraphics[width=7cm]{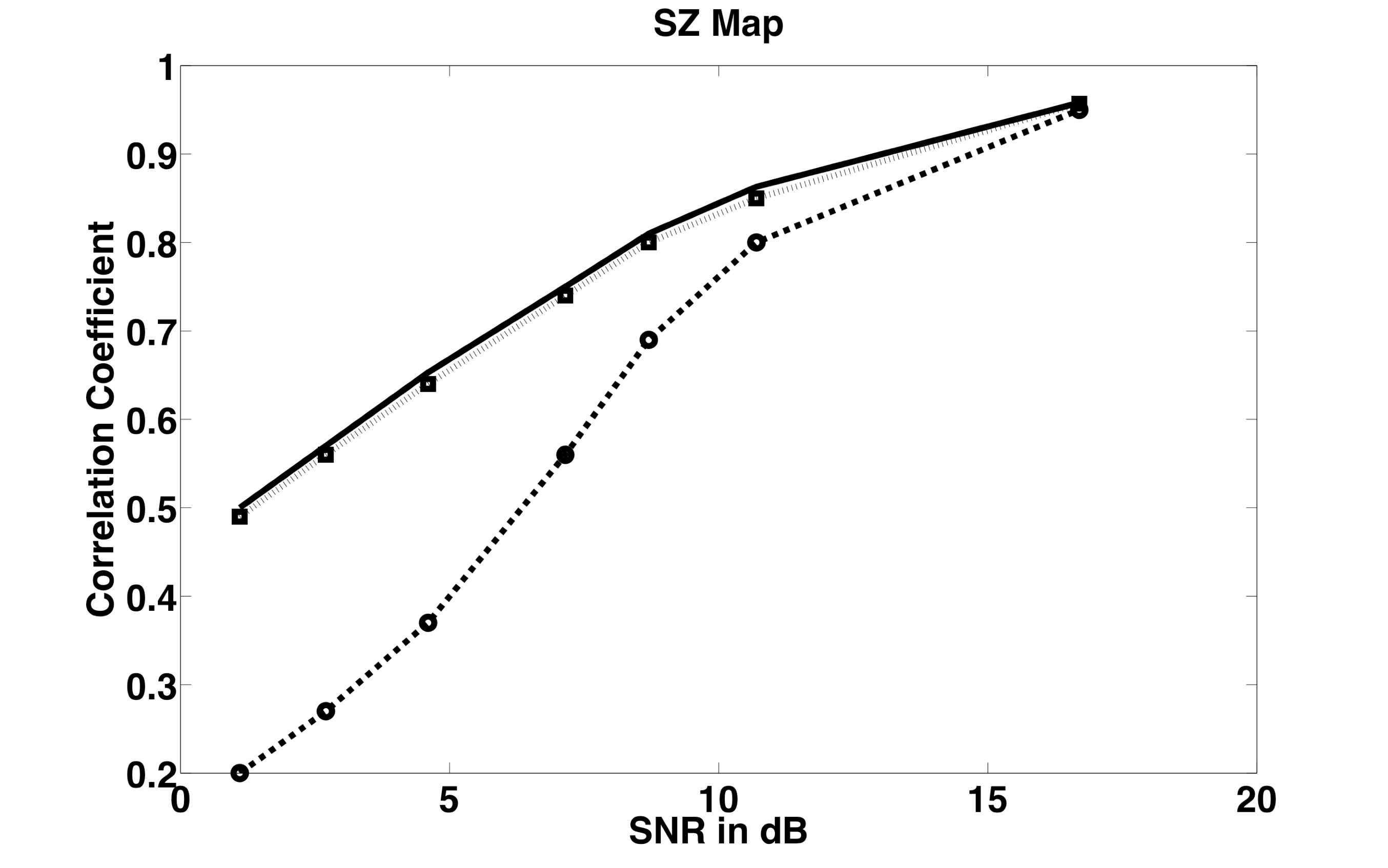}}
\end{minipage}
\hfill
\begin{minipage}[b]{0,5\linewidth}
    \centerline{\includegraphics[width=7cm]{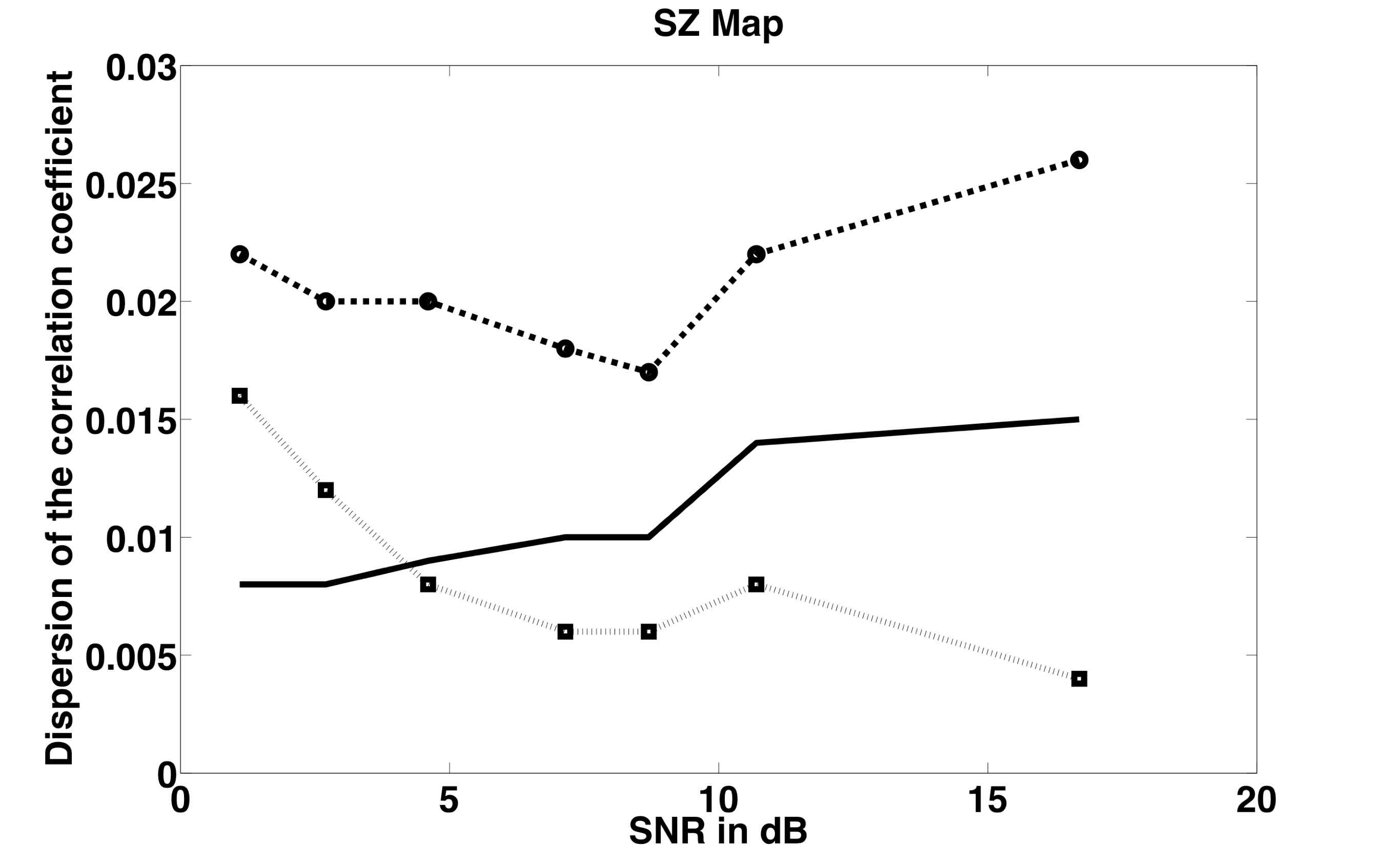}}
\end{minipage}
\vfill
\hspace{0.1in}
\begin{minipage}[b]{0,5\linewidth}
    \centerline{\includegraphics[width=7cm]{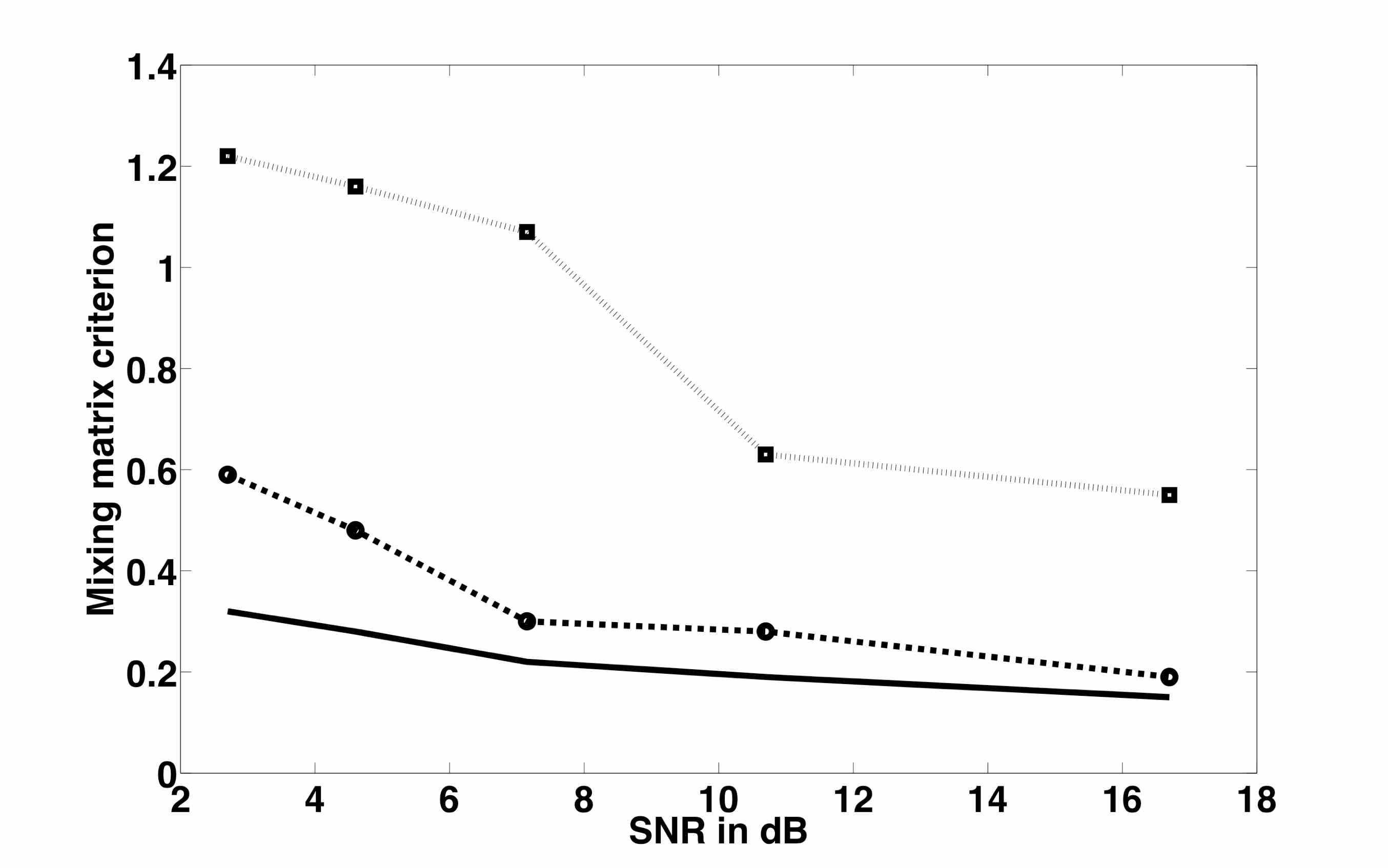}}
\end{minipage}
\hfill
\begin{minipage}[b]{0,5\linewidth}
    \centerline{\includegraphics[width=7cm]{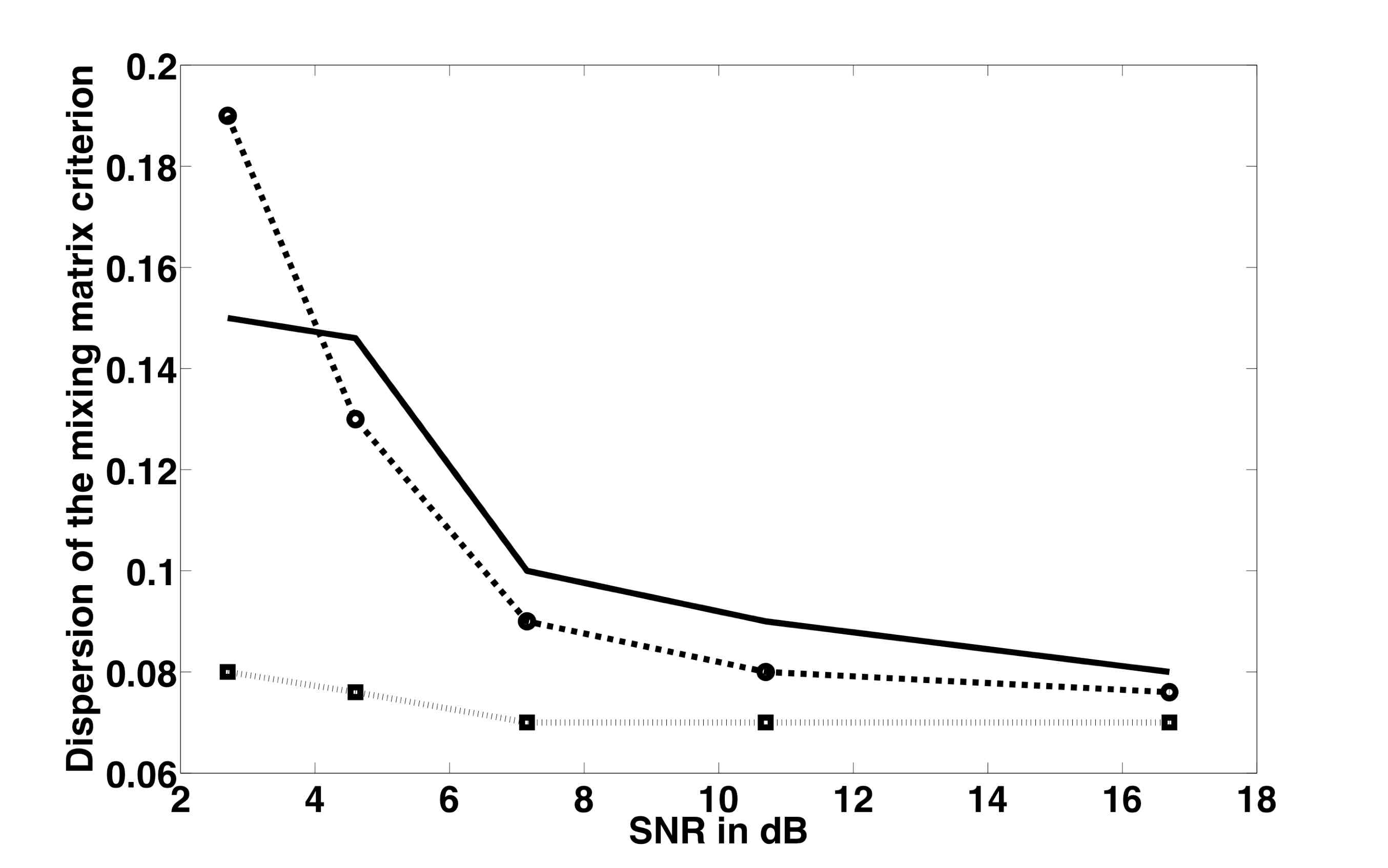}}
\end{minipage}
\vspace{-0.1in} 
\caption{\textbf{Left Column : Mean value of the correlation coefficients between the
estimated source map and the true source map   - Right Column : Dispersion of these correlation coefficients : First line :}  
CMB. \textbf{Second line: } galactic dust. \textbf{Third line: } SZ
map. \textbf{Fourth line: } mixing matrix criterion $\Delta_{\bf A}$ \textbf{Legend :} JADE : \textit{dotted line with }$\Box$ - SMICA
: \textit{dashed line with }$\circ$ - GMCA : \textit{solid
line}. \textbf{Abscissa :} SNR in dB.}
\label{fig:corrsources}
\end{figure}

The method described above was applied to synthetic data composed of
$m=6$ mixtures of $n=3$ sources : CMB, galactic dust emission and SZ maps
illustrated in Figure~\ref{fig:sources} and~\ref{fig:mixtures}. Following \cite{Delab} and \cite{YM}, we
do not take into account the emission at high frequency from the fluctuations of infra-red galaxies.
The synthetic data mimic the observations that will be acquired in the
six frequency channels of Planck-HFI namely : $100,143, 217,353,545$
and $857$ GHz, as shown on Figure~\ref{fig:mixtures}. White Gaussian
noise ${\bf N}$ is added with diagonal covariance matrix ${\bf \Gamma_N}$
reflecting the foreseen Planck-HFI noise levels. Experiments were led
with $7$ \textit{global} noise levels with SNR from $1.7$ to $16.7$dB
such that the experimental noise covariance $\Gamma_{\bf N}$ was
proportional to the nominal noise covariance. Note that the nominal
Planck-HFI global noise level is about $10$dB. Each measurement point
was computed from $30$ experiments involving random noise, randomly
chosen sources from a data set of several simulated CMB, galactic dust
and SZ $256 \times 256$ maps. The astrophysical components and the mixture maps were generated as in~\cite{YM} according to equation (\ref{eq:lm_model}) based on model or experimental emission laws, possibly extrapolated, of the individual components. Separation was obtained with GMCA using
a single orthonormal wavelet basis. Figure~\ref{fig:corrsources} depicts the average correlation coefficients over experiments between the estimated source maps and the true source maps.
Figure~\ref{fig:corrsources} upper left panel shows the
correlation coefficient between the true simulated CMB map and the one
estimated by JADE (\textit{dotted line with }$\Box$), SMICA
(\textit{dashed line with }$\circ$) and GMCA (\textit{solid
line}). The CMB map is well estimated by SMICA, which indeed was
designed for the blind separation of stationary colored Gaussian
processes, but not as well using JADE as one might have expected. GMCA turns out to perform similarly to SMICA. In 
the second line on the left of Figure~\ref{fig:corrsources}, galactic
dust is well estimated by both GMCA and SMICA. The SMICA estimates
seem to have a slightly higher variance than GMCA estimates for higher
global noise levels (SNR lower than $5$ dB). Finally, the picture in the third
line on the left shows that GMCA gives better estimates of the SZ map than SMICA
when the noise variance increases. The right panels provide the dispersion (\emph{i.e.} standard deviation) of the correlation coefficients of the sources estimates. It appears that GMCA is a general
method yielding simultaneous SZ and CMB estimates comparable to state-of-the-art blind separation techniques which seem mostly dedicated to individual components.\\ 
In a noisy context, assessing separation techniques turns out to be more accurate using a mixing matrix criterion. We define
the mixing matrix criterion $\Delta_{\bf A} = \|{\bf I} - {\bf P}{\bf
\hat{A}}^{-1} {\bf A}\|_{1,1}$ (where ${\bf P}$ is a matrix that
reduces the scale/permutation indeterminacy of the mixing model, and
$\|.\|_{1,1}$ is the entrywise $\ell_1$ matrix norm). Indeed, when
${\bf A}$ is perfectly estimated, it is equal to $\bf \hat{A}$ up to
scale and permutation. As we entirely manage our experiments, the true
sources and mixing matrix are known and thus ${\bf P}$ can be computed
easily. The mixing matrix criterion is thus strictly positive unless
the mixing matrix is perfectly estimated up to scale and
permutation. This mixing matrix criterion is experimentally much more
sensitive to separation error. The bottom right panel of Figure~\ref{fig:corrsources} illustrates the
behavior of the mixing matrix criterion $\Delta_{\bf A}$ with JADE,
SMICA and GMCA as the global noise variance varies. GMCA clearly
outperforms SMICA and JADE when applied to CMB data.

\subsection{Adding some physical constraint : the versatility of GMCA}
\label{sec:pbss} 
\begin{figure}[htb]
\vfill
\hspace{0.1in}
\begin{minipage}[b]{0.5\linewidth}
    \centerline{\includegraphics[width=7cm]{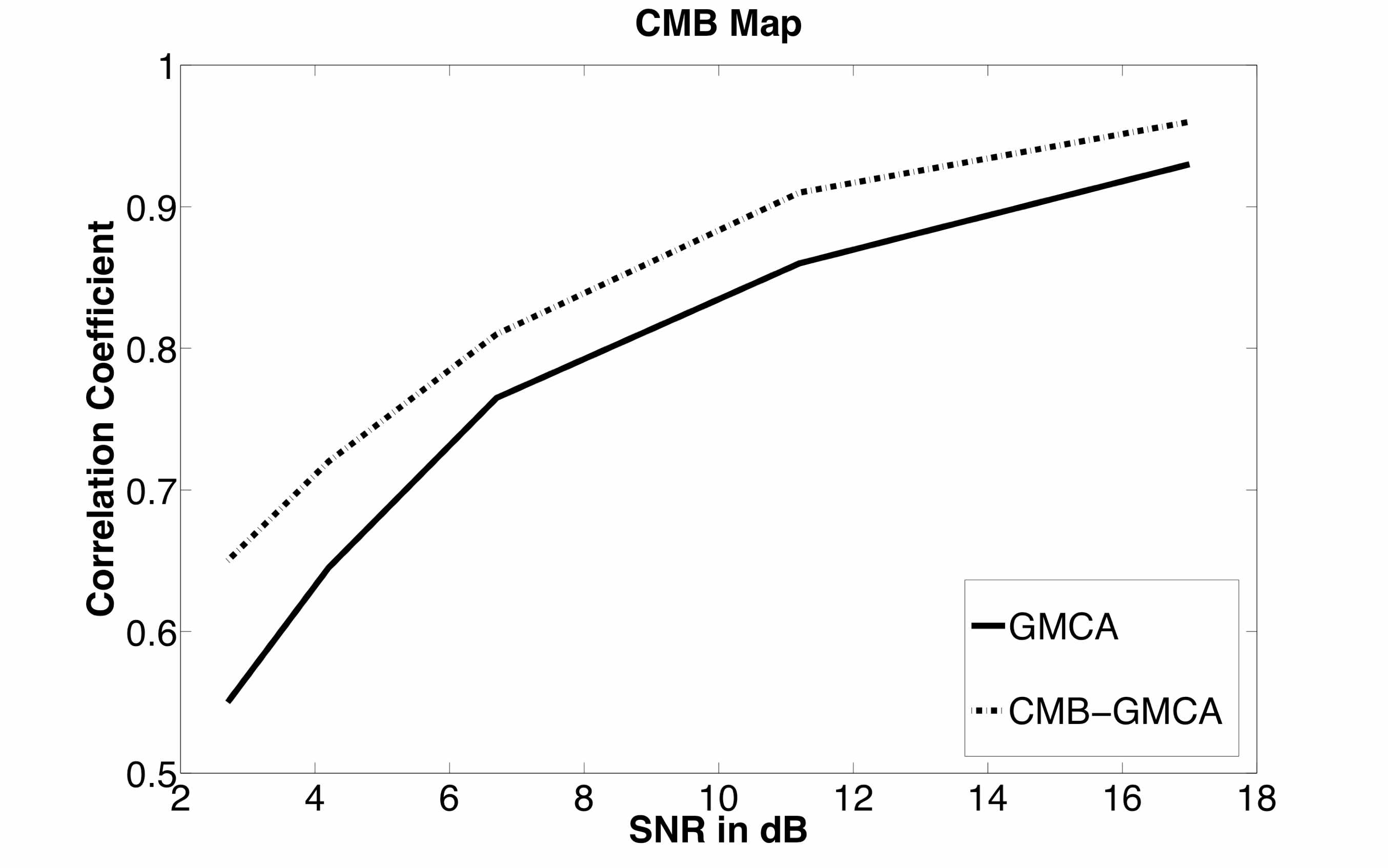}}
\end{minipage}
\hfill
\begin{minipage}[b]{0.5\linewidth}
    \centerline{\includegraphics[width=7cm]{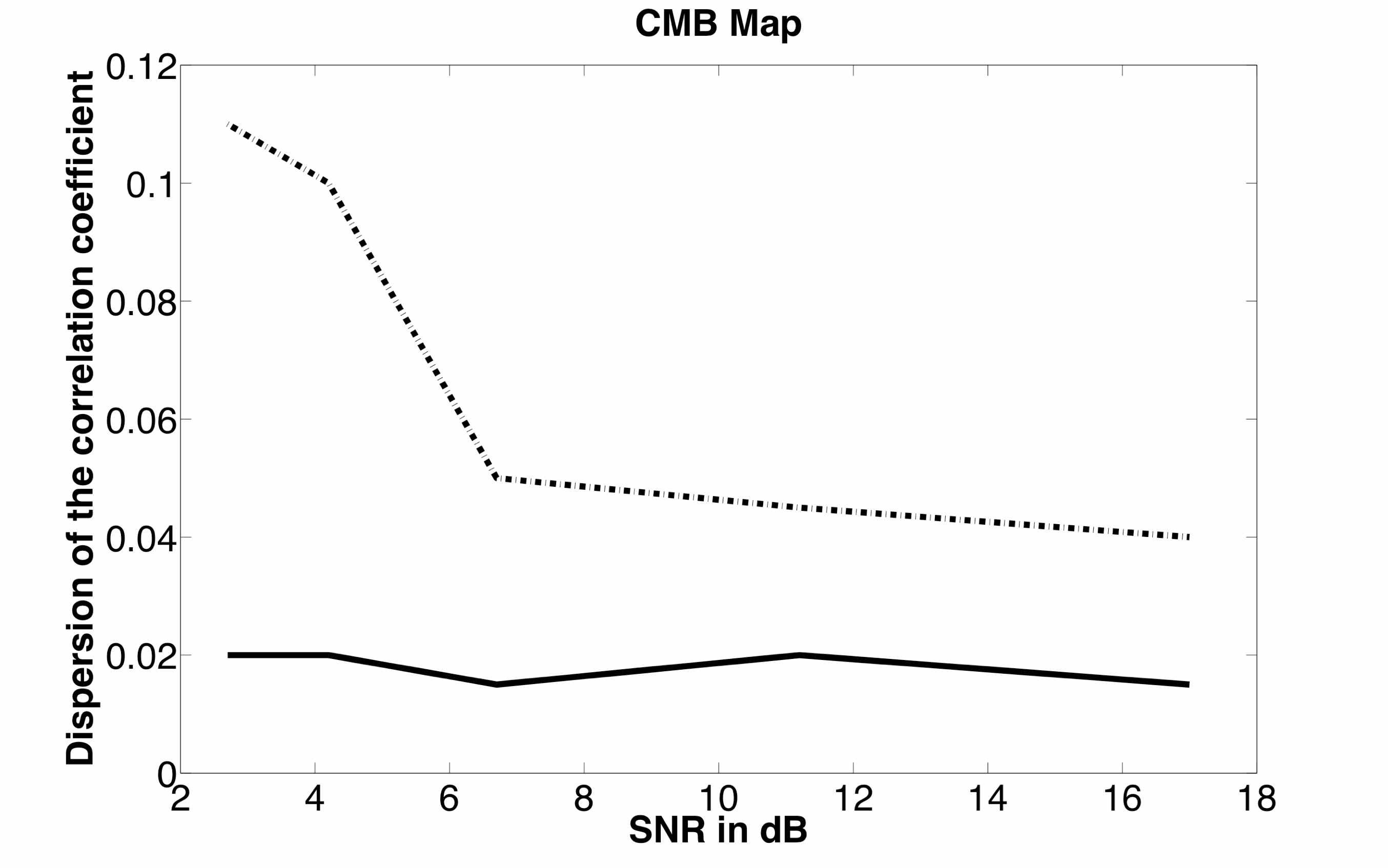}}
\end{minipage}
\vfill
\hspace{0.1in}
\begin{minipage}[b]{0,5\linewidth}
    \centerline{\includegraphics[width=7cm]{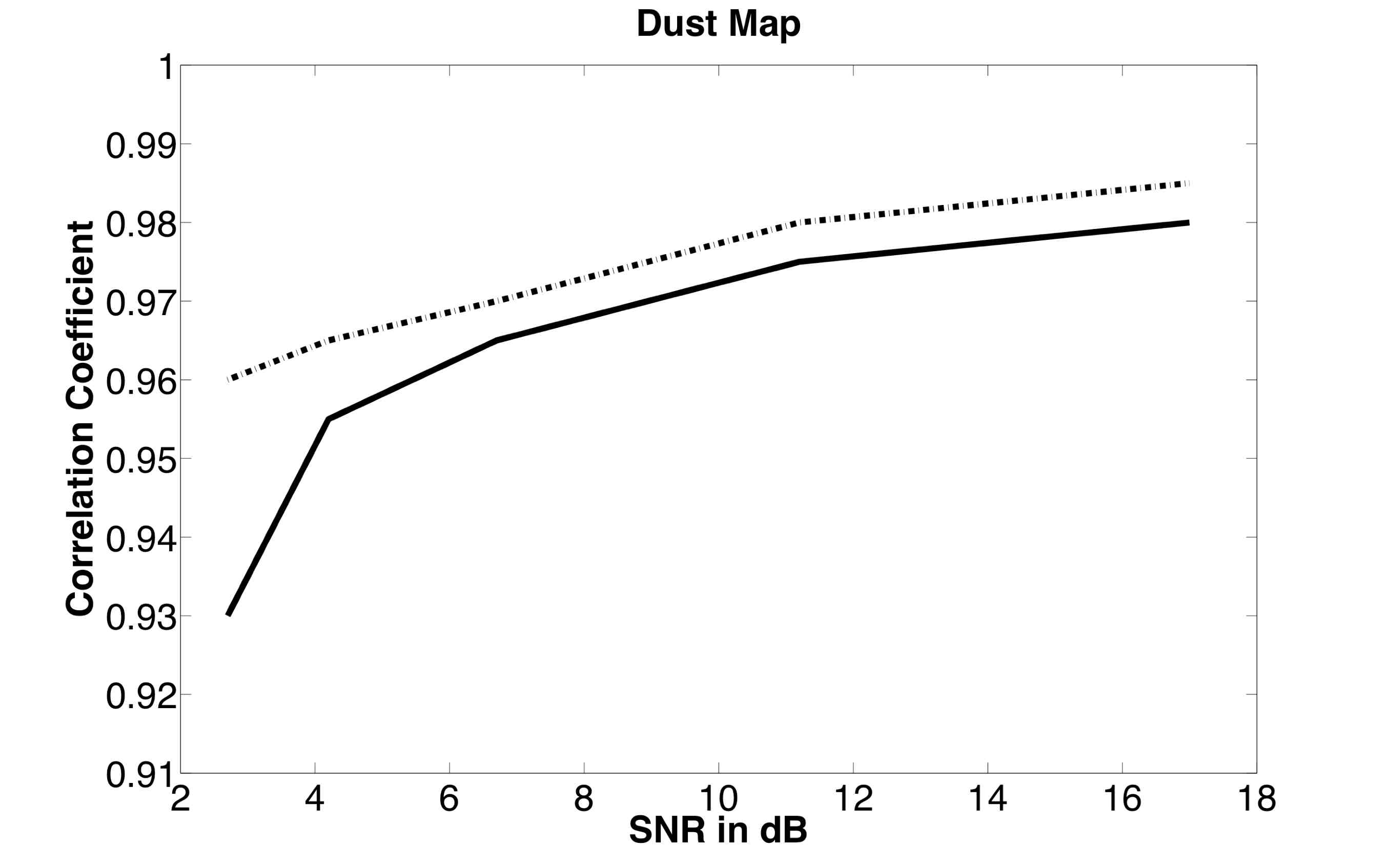}}
\end{minipage}
\hfill
\begin{minipage}[b]{0,5\linewidth}
    \centerline{\includegraphics[width=7cm]{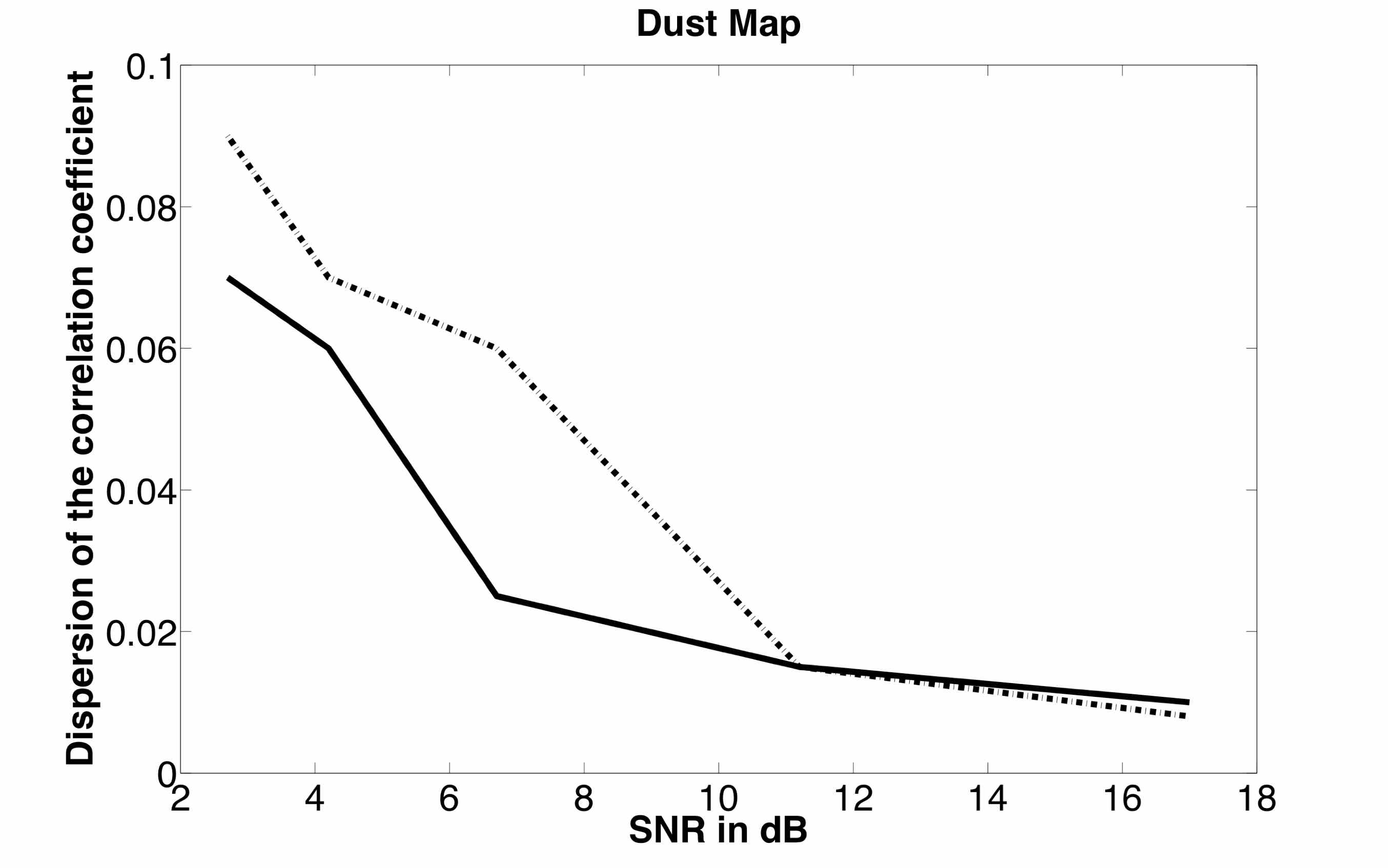}}
\end{minipage}
\vfill
\hspace{0.1in}
\begin{minipage}[b]{0,5\linewidth}
    \centerline{\includegraphics[width=7cm]{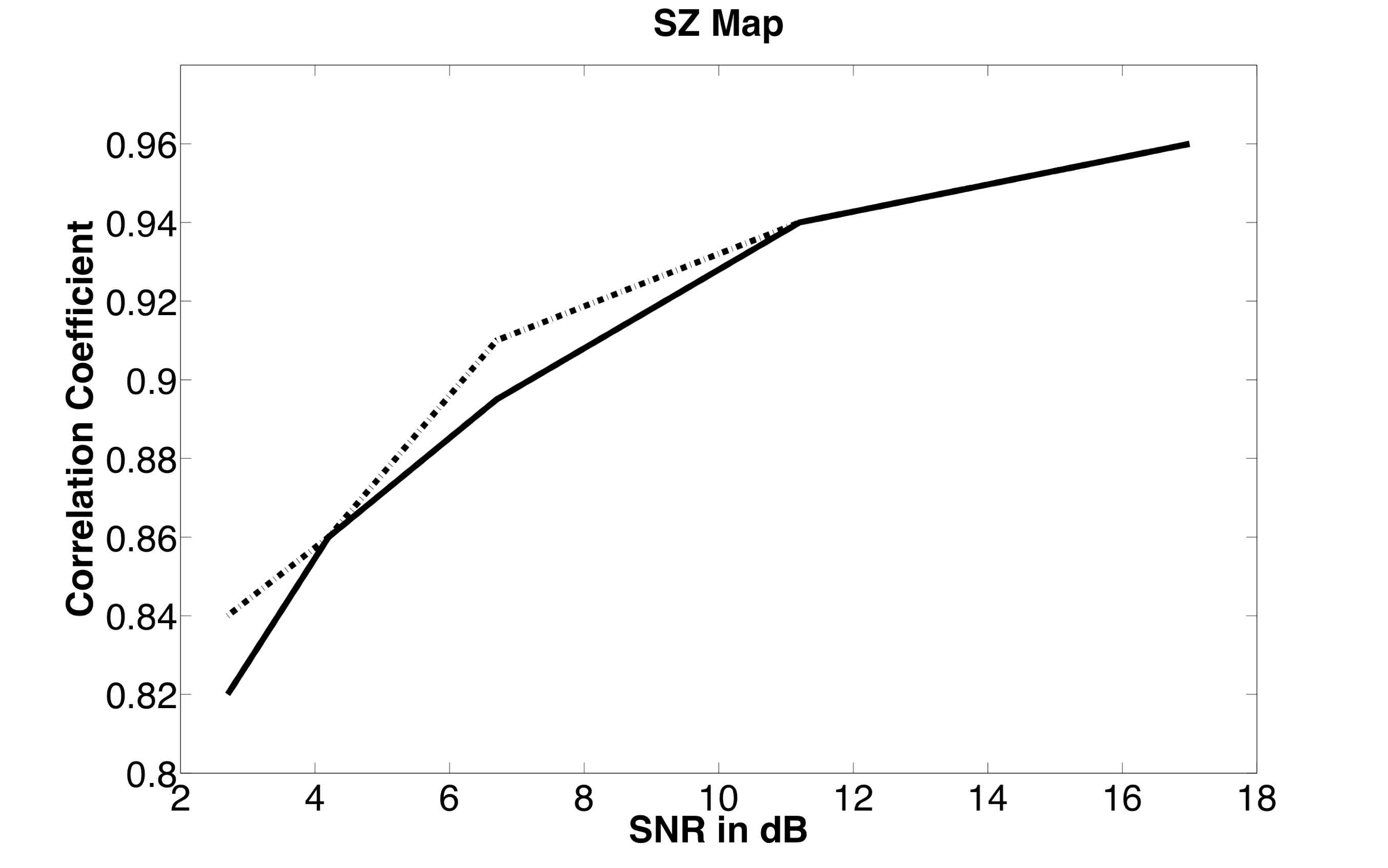}}
\end{minipage}
\hfill
\begin{minipage}[b]{0,5\linewidth}
    \centerline{\includegraphics[width=7cm]{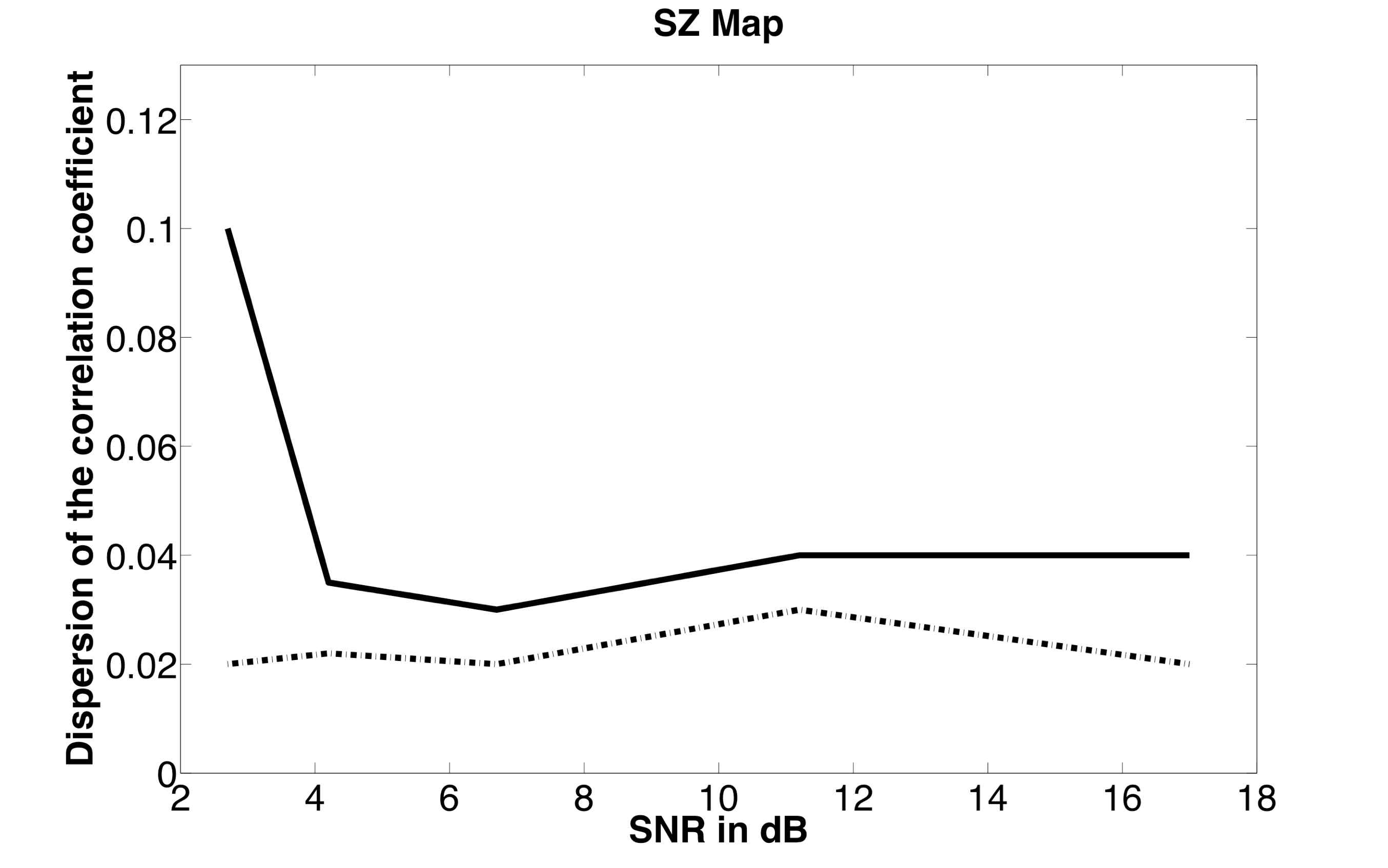}}
\end{minipage}
\vspace{-0.1in} 

\caption{\textbf{Left Column : Mean value of the correlation coefficients between the
estimated source map and the true source map   - Right Column : Dispersion of these correlation coefficients : First line :}  
CMB. \textbf{Second line: } galactic dust. \textbf{Third line: } SZ
map.  \textbf{Legend :} GMCA assuming that the CMB emission law is known : \textit{dotted line} - GMCA : \textit{solid
line}. \textbf{Abscissa :} SNR in dB.}  \label{fig:cmb_sources}
\end{figure}

In practice, the separation task is only partly blind. Indeed, the CMB
emission law is extremely well-known. In this section, we illustrate
that GMCA is versatile enough to account for such prior knowledge. In
the following experiment, CMB-GMCA has been designed by constraining
the column of the mixing matrix ${\bf A}$ related to CMB to its true
value. This is equivalent to placing a strict prior on the CMB column
of ${\bf A}$; that is $P(a^{cmb}) = \delta(a^{cmb} - a^{cmb}_0)$ where
$\delta(.)$ is the Dirac measure and $a^{cmb}_0$ is the true simulated
CMB emission law in the frequency range of
Planck-HFI. Figure~\ref{fig:cmb_sources} shows the correlation
coefficients between the true source maps and the source
maps estimated using GMCA with and without the CMB prior.
As expected, the top left picture of Figure~\ref{fig:cmb_sources}
shows that assuming $a^{cmb}_0$ is known improves the estimation
of CMB. Interestingly, the galactic dust map (in the middle on the left of
Figure~\ref{fig:cmb_sources}) is also better estimated. 
Furthermore, the CMB-GMCA SZ map estimate is likely to have a lower variance (bottom-right panel of
Figure~\ref{fig:cmb_sources}). Moreover, it is likely to provide more
robustness to the SZ and galactic dust estimates thus enhancing the
global separation performances.

\section{Conclusion}
In this paper we underlined that recovering information from CMB data
requires solving a blind source separation issue (BSS). Several BSS
techniques have already been applied to CMB data without providing
good global performances \emph{i.e.} on all components simultaneously. In this paper, we provide a sparsity-based source separation
method coined Generalized Morphological Component Analysis (GMCA)
which turns to give astounding results to effectively recover both CMB
and SZ maps. In that context, sparsity enhances the contrast between
the sources leading to an improved separation task even in a noisy
context. In the blind case, when no prior knowledge is assumed on the
emission laws of the components, GMCA outperforms state-of-the-art
blind component separation techniques already applied to CMB data. Furthermore, GMCA is versatile enough to easily include some
prior knowledge of the emission laws of the components. This is an
extremely valuable feature of the proposed method in the case of CMB
data analysis. Indeed, the CMB has a known black-body
spectrum. Including this information in the GMCA algorithm enhances
the source separation globally. In the present work, we have chosen not to include prior knowledge on the SZ spectral signature. Adding such prior would lead to even better separation results. Future work will be devoted to taking
advantage of GMCA's versatility to adapt to more complex physical
models.

\end{document}